\documentclass[letterpaper,conference]{IEEEtran}
\IEEEoverridecommandlockouts
\usepackage{cite}
\usepackage{amsmath,amssymb,amsfonts}
\usepackage{algorithmic}
\usepackage{graphicx}
\usepackage{textcomp}
\usepackage{xcolor}
\usepackage{array}
\usepackage{dblfloatfix}
\usepackage{subcaption}
\usepackage{multicol}
\usepackage{lipsum}
\usepackage{booktabs}
\usepackage{multirow}
\usepackage{comment}
\usepackage{mathptmx} 
\usepackage{times} 

\def\BibTeX{{\rm B\kern-.05em{\sc i\kern-.025em b}\kern-.08em
    T\kern-.1667em\lower.7ex\hbox{E}\kern-.125emX}}

\begin{document}

\title{G1020: A Benchmark Retinal Fundus Image Dataset for Computer-Aided Glaucoma Detection\\
\thanks{This work is partially funded by National University of Science and Technology (NUST), Pakistan through Prime Minister's Programme for Development of PhDs in Science and Technology, BMBF project DeFuseNN (01IW17002), and NVIDIA AI Lab (NVAIL) programme.}
}

\author{\IEEEauthorblockN{Muhammad Naseer Bajwa}
\IEEEauthorblockA{\textit{Technische Universit\"at Kaiserslautern} \\
\textit{German Research Center for Artificial}\\
\textit{Intelligence GmbH (DFKI)}\\
Kaiserslautern, Germany \\
0000-0002-4821-1056}
\and
\IEEEauthorblockN{Gur Amrit Pal Singh}
\IEEEauthorblockA{\textit{Technische Universit\"at Kaiserslautern} \\
\textit{German Research Center for Artificial}\\
\textit{Intelligence GmbH (DFKI)}\\
Kaiserslautern, Germany \\
0000-0002-6458-9315}
\and
\IEEEauthorblockN{Wolfgang Neumeier}
\IEEEauthorblockA{\textit{Opthalmology Clinic} \\
Kaiserslautern, Germany \\
dr.neumeier-kl@web.de}
\and
\IEEEauthorblockN{Muhammad Imran Malik}
\IEEEauthorblockA{\textit{National University of Science and} \\
\textit{Technology (NUST)} \\
\textit{National Center of Artificial Intelligence}\\
Islamabad, Pakistan \\
0000-0002-8079-5119}
\and
\IEEEauthorblockN{Andreas Dengel}
\IEEEauthorblockA{\textit{Technische Universit\"at Kaiserslautern} \\
\textit{German Research Center for Artificial}\\
\textit{Intelligence GmbH (DFKI)}\\
Kaiserslautern, Germany \\
0000-0002-6100-8255}
\and
\IEEEauthorblockN{Sheraz Ahmed}
\IEEEauthorblockA{\textit{Smart Data and Knowledge Services} \\
\textit{German Research Center for Artificial}\\
\textit{Intelligence GmbH (DFKI)}\\
Kaiserslautern, Germany \\
0000-0002-4239-6520}
}

\maketitle

\begin{abstract}
Scarcity of large publicly available retinal fundus image datasets for automated glaucoma detection has been the bottleneck for successful application of artificial intelligence towards practical Computer-Aided Diagnosis (CAD). A few small datasets that are available for research community usually suffer from impractical image capturing conditions and stringent inclusion criteria. These shortcomings in already limited choice of existing datasets make it challenging to mature a CAD system so that it can perform in real-world environment. In this paper we present a large publicly available retinal fundus image dataset for glaucoma classification called G1020. The dataset is curated by conforming to standard practices in routine ophthalmology and it is expected to serve as standard benchmark dataset for glaucoma detection. This database consists of 1020 high resolution colour fundus images and provides ground truth annotations for glaucoma diagnosis, optic disc and optic cup segmentation, vertical cup-to-disc ratio, size of neuroretinal rim in inferior, superior, nasal and temporal quadrants, and bounding box location for optic disc. We also report baseline results by conducting extensive experiments for automated glaucoma diagnosis and segmentation of optic disc and optic cup.
\end{abstract}


\section{Introduction}
\label{intro}

Computer-Aided Diagnosis (CAD) of ocular diseases is receiving a lot of attention from research community due to its far-reaching benefits of providing swift and accurate large-scale screening as well as reducing physicians' workload in routine clinical setup~\cite{hagiwara2018computer}. Machine Leaning (ML) and Deep Learning (DL) based techniques are commonly used to automatically detect various ocular diseases like glaucoma~\cite{rogers2019evaluation}, diabetic retinopathy~\cite{bajwa2019combining}, Age-related Macular Degeneration (AMD)~\cite{pead2019automated} and many other retinal disorders~\cite{son2020development}. Recently, it has been shown that Retinal Fundus Images (RFIs) can be used to detect many non-ocular diseases as well like Type-II diabetics~\cite{heslinga2019direct}, anaemia~\cite{mitani2019detection}, and cardiovascular risks~\cite{poplin2018prediction}. For automated glaucoma detection, different image modalities and clinical tests are used, for instance, RFIs~\cite{bajwa2019two}, Optical Coherence Tomography (OCT)~\cite{an2019glaucoma}, and Visual Field Tests (VFTs)~\cite{kucur2018deep}. However, fundus imaging is the most common and inexpensive imaging technique \cite{ABRAMOFF2013151} for large-scale screening of various retinal diseases.

Most of the publicly available RFI datasets have only a few hundred images (see section \ref{relatedWork}). These datasets are collected with many imaging constraints like centralising Optic Disc (OD)~\cite{zhang2010origa} or macula and removing images containing certain artefacts~\cite{diaz2019cnns}. Since the most important application of automated glaucoma detection is cost-effective and large-scale screening \cite{li2018efficacy} of general population, these automated solutions should be able to perform well in real-world scenarios with fundus images taken in day-to-day practice without many constraints \cite{hood2018efficacy}. Removing images that do not conform to strict inclusion criteria for example, from the available datasets might result in a CAD that works exceptionally well in controlled \emph{laboratory} environment but might fail in routine screening or clinical workflow.

In this paper we present a new publicly available RFI dataset called G1020\footnote{Available at: https://www.dfki.uni-kl.de/g1020} for segmentation of OD and Optic Cup (OC) and detection of glaucoma. This dataset contains images taken under realistic conditions without many imaging constraints and, as a result, is fairly representative of real-world fundus imaging practices. We provide ground truth annotations for OD and OC segmentation, bounding box coordinates for OD localisation, vertical Cup-to-Disc Ratio (CDR), and size of neuroretinal rim in Inferior, Superior, Nasal and Temporal quadrants to see if ISNT rule is followed. We also provide gold standard clinical diagnosis for glaucoma and many other ocular disorders. We believe that this challenging dataset can be used as a benchmark dataset to train robust algorithms for glaucoma detection capable of performing in the field or in clinics.

\section{Related Work}
\label{relatedWork}
In this section we first present some of the largest publicly available RFI datasets for glaucoma detection and segmentation of OD and OC. Later, we survey a handful of contemporary works involving segmentation and classification tasks using these and other datasets.

\subsection{Existing RFI Datasets}
\subsubsection{ORIGA} Online Retinal fundus Image database for Glaucoma Analysis and research (ORIGA)~\cite{zhang2010origa} is one of the largest and most commonly used dataset for glaucoma detection made public since 2010. This dataset consists of 650 images (168 glaucoma, 482 healthy) collected by Singapore Eye Research Institute between 2004 and 2007. The dataset provides class labels for healthy and glaucoma, OD and OC contours and CDR values for each image.

\subsubsection{RIM-ONE} This small dataset~\cite{fumero2011rim-one} consists of 169 high resolution RFIs collected at three Spanish hospitals. Each image is classified as healthy, early glaucoma, moderate glaucoma, deep glaucoma or ocular hypertension. Additionally, it provides OD segmentation annotations to evaluate OD detection algorithms.

\subsubsection{RIGA} Retinal fundus Images for Glaucoma Analysis (REGA)~\cite{almazroa2018RIGA} consists of 750 images taken from Messidor dataset~\cite{decenciere2014Messidor} and two clinic in Saudi Arabia. This dataset provides OD and OC boundary annotations; however, it does not provide any diagnosis with regards to glaucoma.

\subsubsection{REFUGE} REtinal FUndus Glaucoma ChalengE (RIGA)~\cite{orlando2020refuge} is the largest and one of the latest RFI datasets publicly available for glaucoma detection. It was made public in 2018 as a grand challenge and consists of 1200 fundus images with ground truth segmentation of OD and OC and clinical glaucoma labels. Despite large size of this dataset, this dataset is highly unbalanced towards healthy class as it contains only 120 glaucoma images.

\subsubsection{ACRIMA} This new dataset~\cite{diaz2019cnns} consists of a total of 705 fundus images with 396 glaucoma images and 309 normal images taken with centred optic disc. The dataset does not provide any annotations for OD and OC segmentation. Relatively balanced proportion of normal and glaucomatous images in this dataset makes it particularly suitable for training DL based classifiers.

\subsection{Optic Disc and Optic Cup Segmentation}
Almazroa et al. \cite{almazroa2017optic} devised an image processing based heuristic algorithm for optic disc segmentation using RIGA dataset, which was later made public \cite{almazroa2018RIGA}. Their algorithm achieved an accuracy of 83.9\% for marking the OD area and centroid. Al-Bander et al.~\cite{al2018dense} used a U-Net~\cite{ronneberger2015unet} like dense fully connected Convolutional Neural Network (CNN) for OD and OC segmentation and evaluated their method on 1129 RFIs from five public datasets. Their method was shown to be invariant to population demography, camera models, and other ocular diseases. They outperformed the state-of-the-art on two datasets and gave competitive results on two datasets without training on these four datasets. Fu et al.~\cite{fu2018joint} attempted to jointly segment OD and OC. They modified faster R-CNN~\cite{ren2015faster} by replacing its Region Proposal Network (RPN) with two networks named Disc Proposal Network (DPN) and Cup Proposal Network (CPN). They tested their proposed network on publicly available ORIGA dataset and 1676 image of  a private dataset called SCES~\cite{baskaran2015prevalence}, and outperformed state-of-the-art methods for joint segmentation of OD and OC.

\subsection{Glaucoma Classification}
Raghavendra et al. \cite{raghavendra2018deep} used 1426 private RFIs to train and test an 18-layer Deep Neural Network (DNN) and achieved 95.6\% accuracy, 95.5\% sensitivity and 95.7\% specificity for glaucoma classification. In a large and comprehensive study using around 40,000 RFIs, Li et al. \cite{li2018efficacy} evaluated the performance of inception v3 for detecting referable Glaucomatous OpticNeuropathy (GON). They defined GON as vertical CDR greater than 0.7. They achieved 92.9\% accuracy and 98.6\% Area Under the Curve (AUC) with 95.6\% sensitivity and 92.0\% specificity. They found that the leading reason for false positive results was presence of other eye conditions in the fundus images. Al-Bander et al.~\cite{al2017automated} used 455 images of RIM-ONE v2 dataset and extracted discriminating features using DNN before classifying them using Support Vector Machine (SVM). They obtained 88.2\% accuracy, 85\% sensitivity and 90.8\% specificity.

\section{Dataset Description}
\begin{figure*}[h!]%
\centering
\begin{subfigure}{0.89\columnwidth}
\includegraphics[width=\columnwidth]{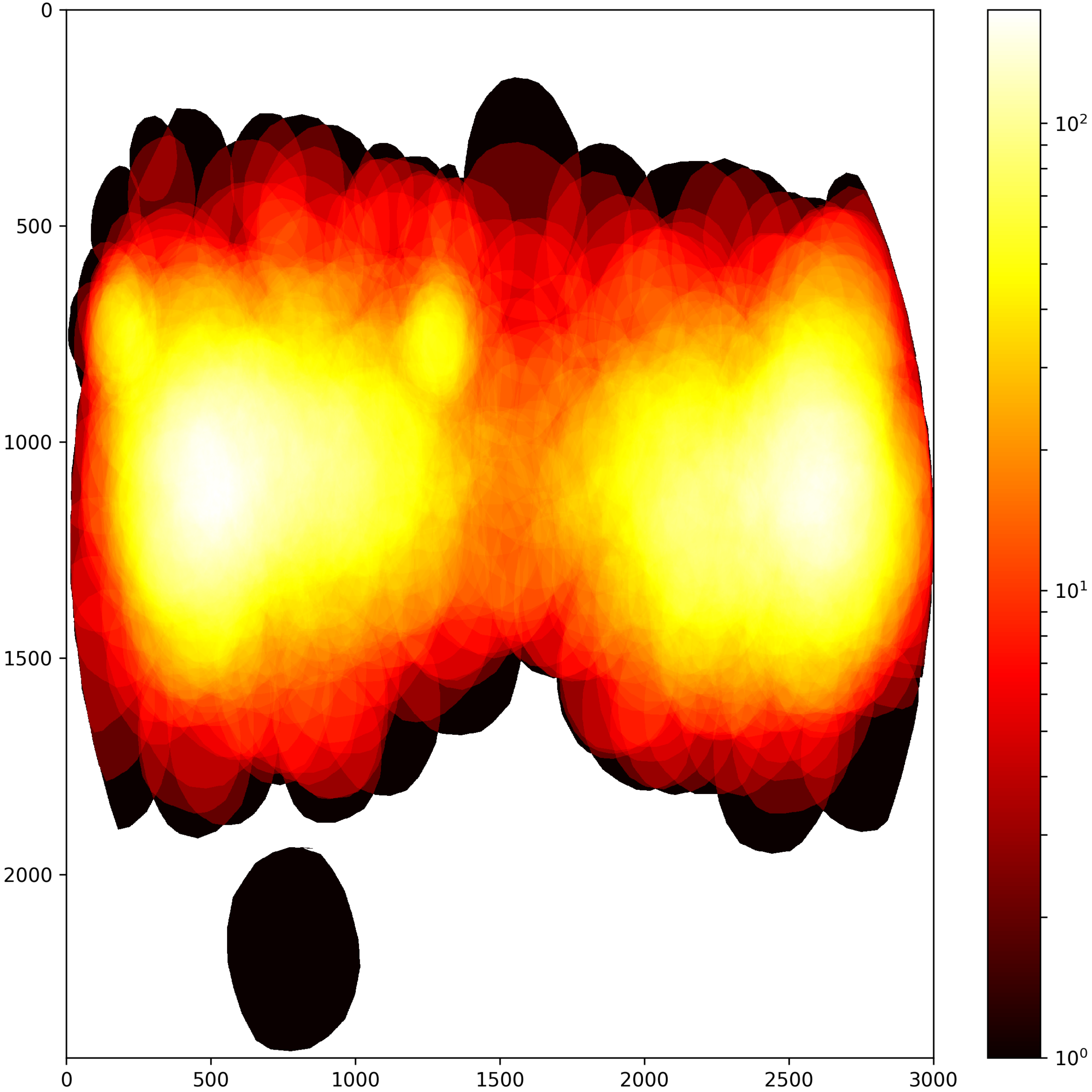}%
\caption{G1020}%
\label{subfig:densityG1020}%
\end{subfigure}
\hfill%
\begin{subfigure}{0.89\columnwidth}
\includegraphics[width=\columnwidth]{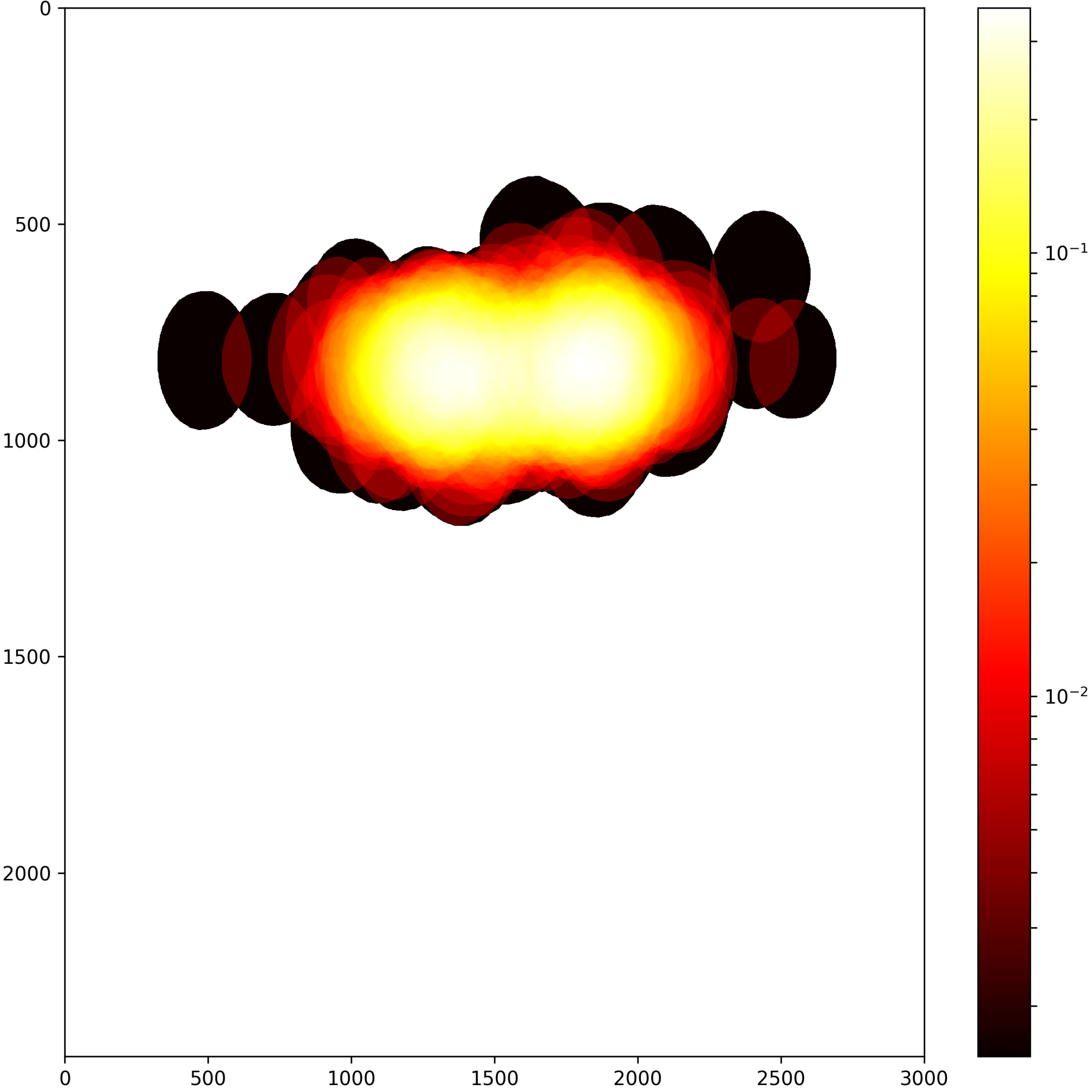}%
\caption{ORIGA}%
\label{subfig:densityORIGA}%
\end{subfigure}
\caption{Density Map of optic disc in G1020 and ORIGA. Optic disc in G1020 is not centralised, making post-processing of segmentation algorithms more challenging.}
\label{fig:densityMaps}
\end{figure*}

\begin{figure*}[b]%
\centering
\begin{subfigure}{0.89\columnwidth}
\includegraphics[width=\columnwidth]{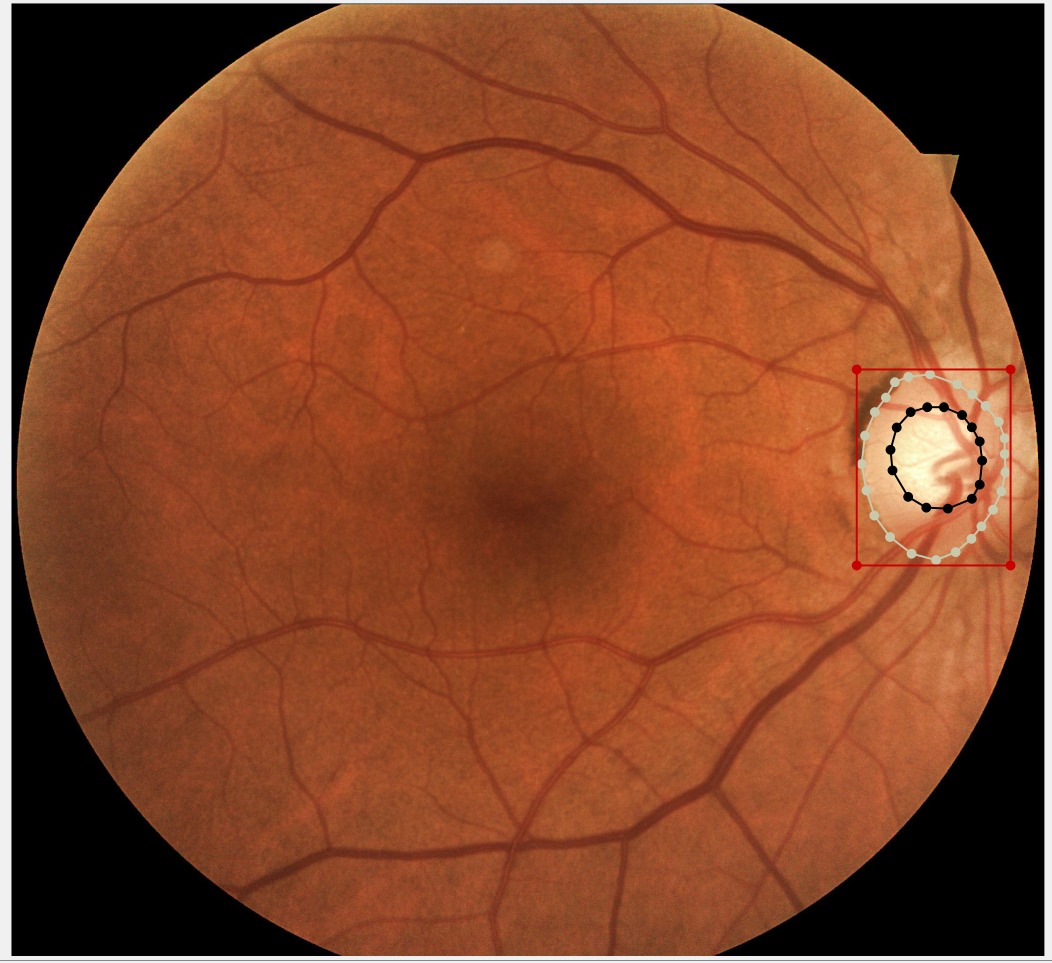}%
\caption{Sample image with all three annotations}%
\label{subfig:cupNdisc}%
\end{subfigure}
\hfill%
\begin{subfigure}{0.89\columnwidth}
\includegraphics[width=\columnwidth]{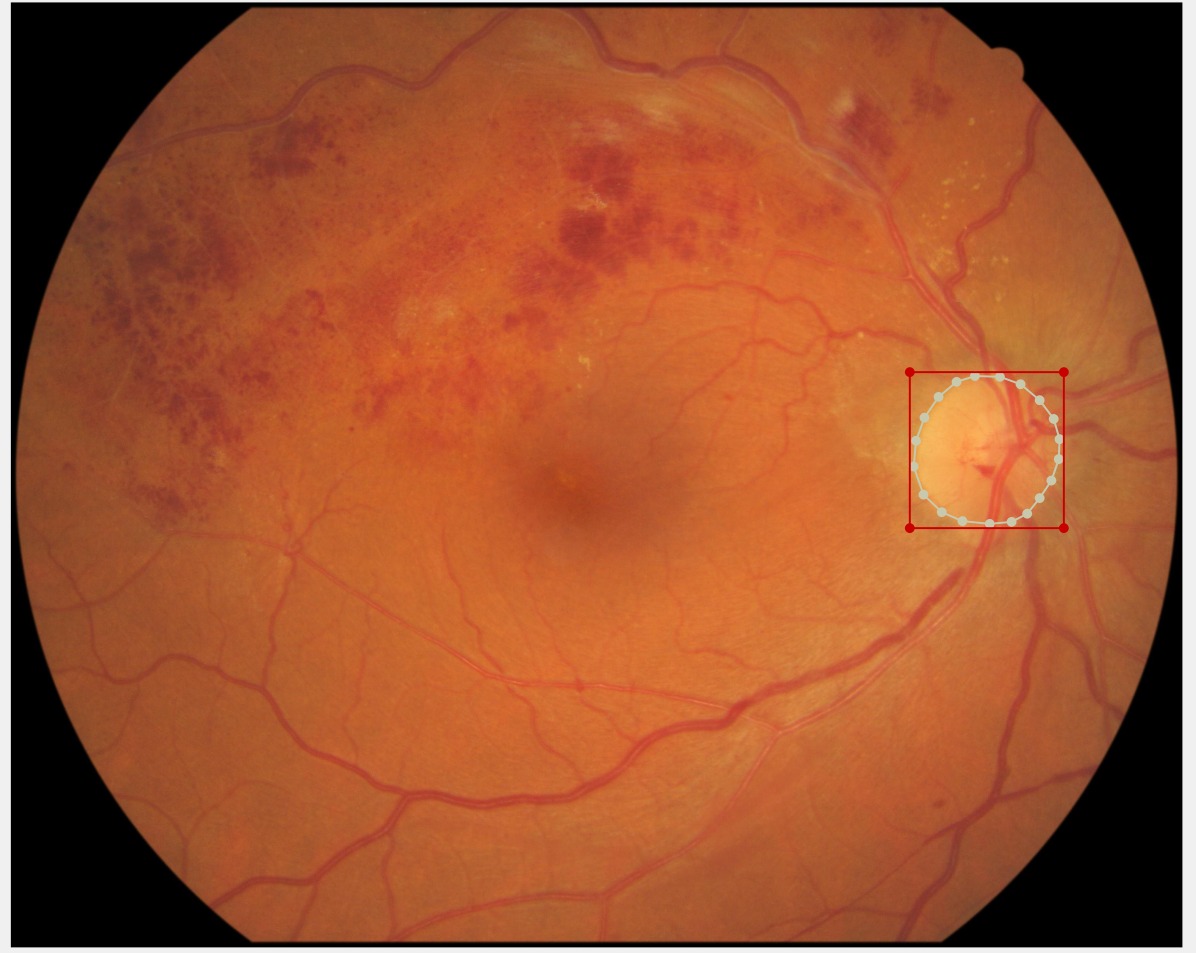}%
\caption{Sample image without optic cup}%
\label{subfig:discOnly}%
\end{subfigure}
\caption{Sample images with optic cup (black polygon), optic disc (white polygon) and bounding box (red rectangle) annotations.}
\label{fig:annot}
\end{figure*}

\begin{figure*}[b!]%
\centering
\begin{subfigure}{0.89\columnwidth}
\includegraphics[width=\columnwidth]{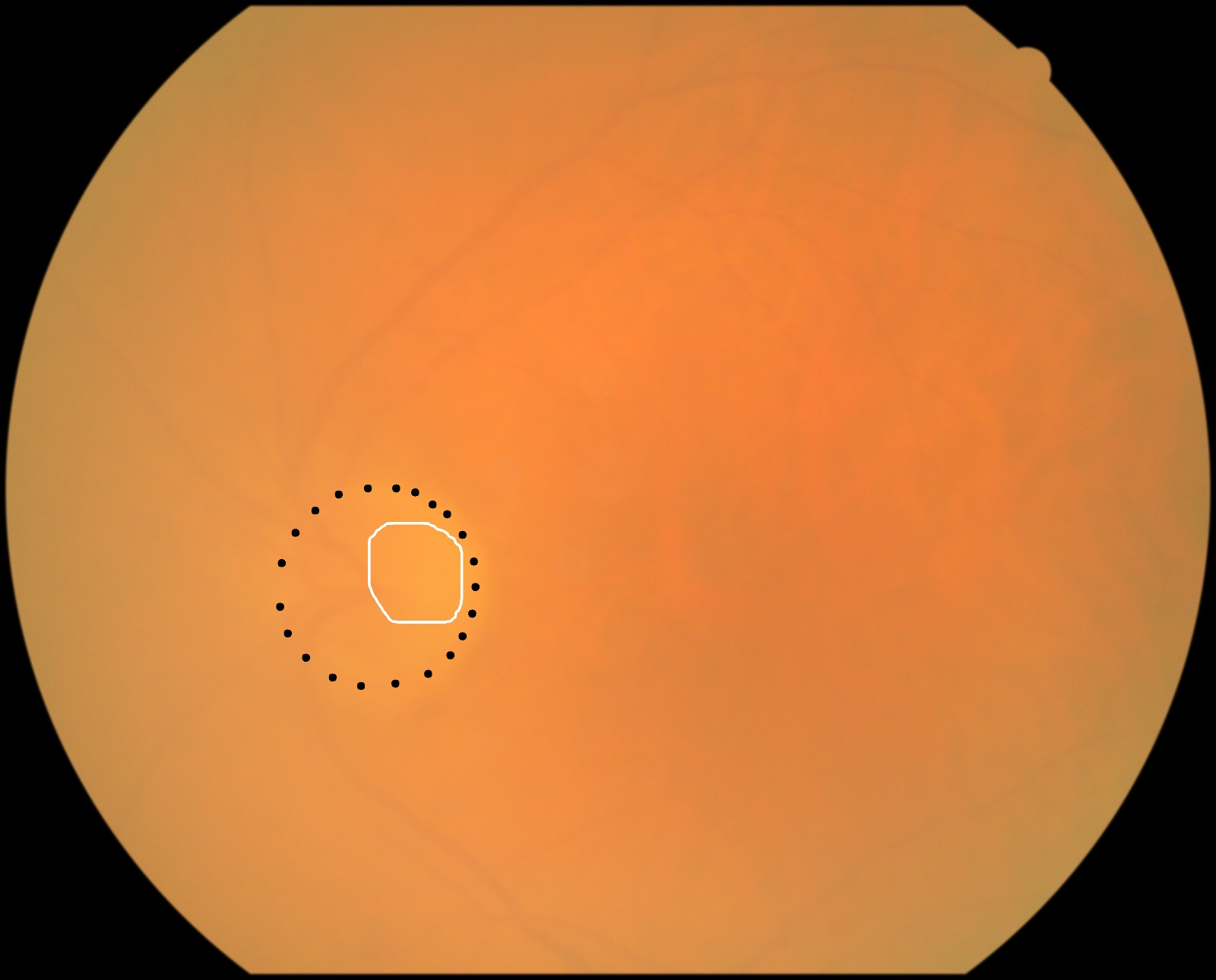}%
\caption{Image with least IOU (= 0.2689) between prediction and GT of OD}%
\label{subfig:missOD}%
\end{subfigure}
\hfill%
\begin{subfigure}{0.89\columnwidth}
\includegraphics[width=\columnwidth]{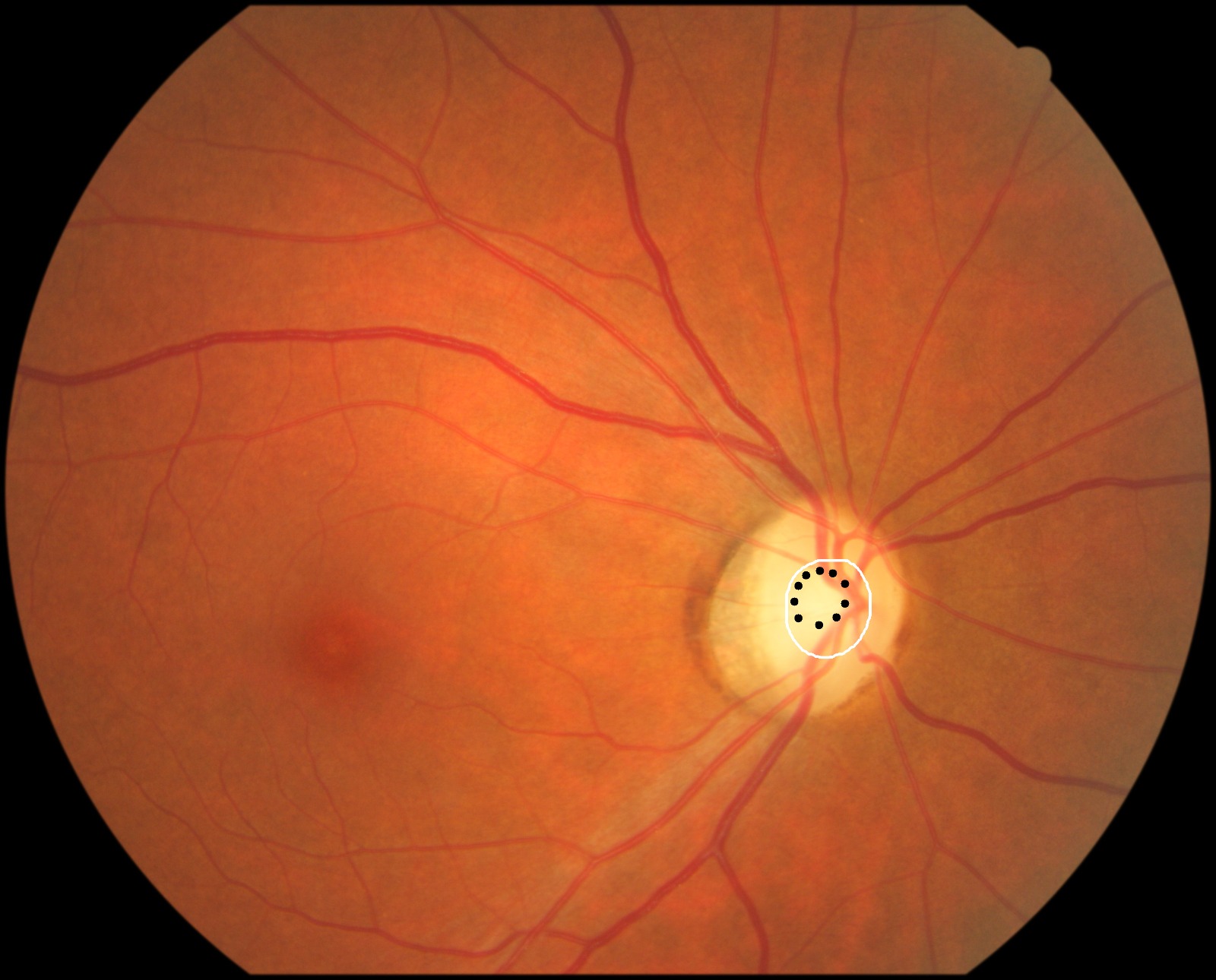}%
\caption{Image with least IOU (= 0.308) between prediction and GT of OC}%
\label{subfig:missOC}%
\end{subfigure}
\caption{Example images with incorrect OD and OC detection. Dotted annotations correspond to GT, whereas solid annotations represent prediction.}
\label{fig:missDdetect}
\end{figure*}
The images in G1020 are collected at a private clinical practice in Kaiserslautern, Germany between year 2005 and 2017 with 45-degree field of view after using dilation drops. The records were subsequently anonymised and random unique patient identifiers were assigned to each record. Because the images are collected retrospectively and are fully anonymised the informed consent of the patients was not required. To achieve a dataset that reflects routine clinical practice at busy healthcare facilities, no specific imaging constraints, like centring of OD or macula, were imposed. Fig.~\ref{fig:densityMaps} shows density map of OD in all images of G1020 as compared to corresponding density map of ORIGA. It can be seen that images in G1020 dataset have OD at a wider spatial area making post-processing of any segmentation algorithm significantly challenging. The images are stored in .JPG format. In the final dataset released, black background is truncated and only the fundus region is preserved resulting in images of size between $1944\times2108$ and $2426\times3007$ pixels.

There are total of 1020 images from 432 patients. Each patient has a minimum of 1 image and maximum of 12 images. Out of 1020 images, 296 images from 110 patients were found to have glaucoma and 724 images from 322 patients were healthy. There was no patient with images belonging to both healthy and glaucomatous class.

\begin{table*}[h!]
\caption{Segmentation performance of Mask R-CNN on G1020 dataset.}
\label{tab:segResults}
\centering
\begin{tabular}{@{}ccccccc@{}}
\toprule
\multicolumn{1}{c}{\textbf{Train/Test Splits}} & \textbf{Object} & \textbf{Criterion} & \textbf{Average IOU} & \textbf{Precision} & \textbf{Recall} & \textbf{F1-Score} \\ \midrule
\multirow{6}{*}{\begin{tabular}[c]{@{}l@{}}Train: G1020\\ (random 80\%)\\ Test: G1020\\ (random 20\%)\end{tabular}} & \multirow{3}{*}{Optic Disc} & IOU\textgreater{}0.4 & 0.8852 & 0.9951 & 0.9951 & 0.9951  \\
 &  & IOU\textgreater{}0.5 & 0.8852 & 0.9951 & 0.9951 & 0.9951  \\
 &  & IOU\textgreater{}0.6 & 0.8852 & 0.9951 & 0.9951 & 0.9951  \\ \cmidrule(l){2-7} 
 & \multirow{3}{*}{Optic Cup} & IOU\textgreater{}0.4 & 0.7276 & 0.9810 & 0.9810 & 0.9810  \\
 &  & IOU\textgreater{}0.5 & 0.7364 & 0.9494 & 0.9494 & 0.9494  \\
 &  & IOU\textgreater{}0.6 & 0.7645 & 0.8228 & 0.8228 & 0.8228  \\ \midrule
\multirow{6}{*}{\begin{tabular}[c]{@{}l@{}}Train: ORIGA\\ (all images)\\ Test: G1020\\ (all images)\end{tabular}} & \multirow{3}{*}{Optic Disc} & IOU\textgreater{}0.4 & 0.8641 & 0.9920 & 0.9774 & 0.9847  \\
 &  & IOU\textgreater{}0.5 & 0.8665 & 0.9861 & 0.9716 & 0.9786 \\
 &  & IOU\textgreater{}0.6 & 0.8719 & 0.9692 & 0.9549 & 0.962  \\ \cmidrule(l){2-7} 
 & \multirow{3}{*}{Optic Cup} & IOU\textgreater{}0.4 & 0.6496 & 0.9071 & 0.9014 & 0.9042  \\
 &  & IOU\textgreater{}0.5 & 0.6809 & 0.7812 & 0.7762 & 0.7787  \\
 &  & IOU\textgreater{}0.6 & 0.7256 & 0.5489 & 0.5752 & 0.5770  \\ \bottomrule
\end{tabular}
\end{table*}

Clinical diagnosis is provided for each patient with regards to presence or absence of glaucoma and any other ocular disorder observed. To provide segmentation ground truth, an expert marked OD and OC boundaries as well as bounding box annotations using \emph{labelme} \cite{russell2008labelme}, which is an open source annotation tool developed by MIT. These manual annotations are verified and corrected (if required) by a veteran ophthalmologist with more than 25 years of clinical experience. The annotations are saved in JSON files corresponding to each image. Based on the ground truth annotations for OD and OC, vertical CDR is calculated and size of neuroretinal rim in four quadrants is measured to see if ISNT rule is followed. In 60 glaucomatous images, OC was not visible whereas 170 healthy images also do not show any visible OC. Fig. \ref{fig:annot} shows sample images with OD, OC, and bounding box annotations.
\begin{figure*}[b!]%
\centering
\begin{subfigure}{\columnwidth}
\includegraphics[width=\columnwidth]{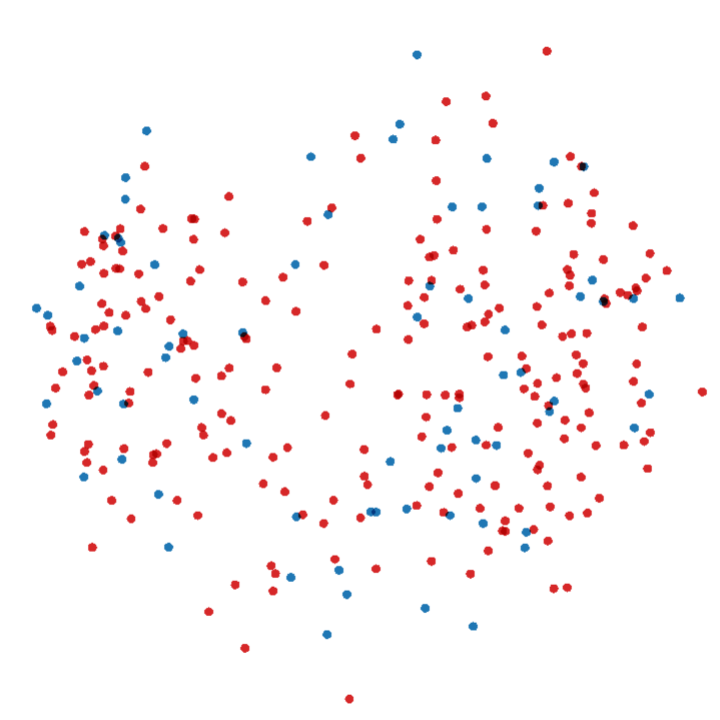}%
\caption{G1020}%
\label{subfig:pcaG1020}%
\end{subfigure}
\hfill%
\begin{subfigure}{\columnwidth}
\includegraphics[width=\columnwidth]{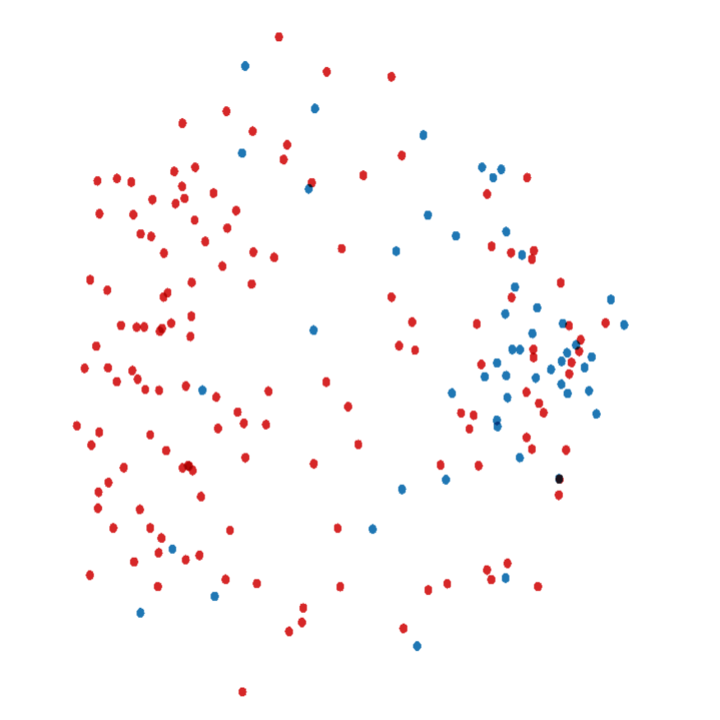}%
\caption{ORIGA}%
\label{subfig:pcaORIGA}%
\end{subfigure}
\caption{Visualisation of image embeddings on 2D plane after dimensionality reduction using PCA for G1020 and ORIGA. Blue dots represent glaucoma images and red dots represent healthy images.}
\label{fig:PCA}
\end{figure*}

\section{Experiments and Evaluation Results}

We evaluated state-of-the-art segmentation algorithms and image classification networks on our G1020 dataset. For automated segmentation of OD and OC we used Mask R-CNN \cite{he2017mask} with ResNet-50 \cite{he2016resnet} as convolutional backbone pre-trained on ImageNet \cite{krizhevsky2012imagenet}. We trained separate models for segmentation of OD and OC. We first trained using 80\% random images from G1020 and tested on remaining 20\% images. The names of images in both training and testing splits are given with the dataset. Secondly we trained Mask-RCNN using all images of ORIGA and evaluated their performance on all images of G1020. Table \ref{tab:segResults} summarises segmentation results. We employed multiple criteria to consider a detected OD and OC as correct or incorrect. Table \ref{tab:segResults} shows results for three such criteria, namely when Intersection Over Union (IOU) between predicted object and ground truth object is $> 40, 50$ or $60$.

To refine our segmentation results, we employed Non-Maximum Suppression (NMS) and got rid of all but one contour with highest probability score. If the overlap (IOU) between a predicted object (OD or OC) and it's ground truth is less than the criterion (IOU $>0.4$, for example), it's considered as both a False Negative (FN), since the actual object is not detected, and a False Positive (FP), since an object other than actual object is predicted. For training and testing on G1020 the network was able to predict OC and OD for each image. In this experiment there was only one image with IOU = 0.2689 below three criteria given in Table~\ref{tab:segResults}. Second minimum IOU  was found to be 0.6429. Therefore, precision, recall, and F-1 score for all three criteria are the same. Furthermore, since the only misclassified image resulted in 1 FP and 1 FN, therefore, the values of precision and recall are also the same. For experiment with training using ORIGA and testing on G1020, the network was able to detect 786 cups out of 791 actual cups and 1005 discs out of 1020 discs. Therefore, precision and recall are different in that experiment for each criterion. Fig.~\ref{fig:missDdetect} shows sample images with incorrectly detected OD and OC. 

\begin{table}[t!]
\centering
\caption{Mean Absolute Percentage Error (MAPE) of various parameters for correctly detected optic disc and optic cup. STD stands for Standard Deviation.}
\label{tab:MAPE}
\begin{tabular}{@{}ccccc@{}}
\toprule
\textbf{Train/Test Split} & \multicolumn{2}{c}{\textbf{Parameters}} & \textbf{Mean} & \textbf{STD} \\ \midrule
\multirow{7}{*}{\begin{tabular}[c]{@{}c@{}}Train: G1020\\ (random 80\%)\\ Test: G1020\\ (random 20\%)\end{tabular}} & \multicolumn{2}{l}{Cup Diameter} & 0.2242 & 0.1933 \\
 & \multicolumn{2}{l}{Disc Diameter} & 0.0502 & 0.0664 \\
 & \multicolumn{2}{l}{CDR} & 0.2304 & 0.1852 \\
 & \multirow{4}{*}{\begin{tabular}[l]{@{}l@{}}Neuroretinal\\ Rim\end{tabular}} & Inferior & 0.1226 & 0.1002 \\
 &  & Superior & 0.0206 & 0.0314 \\
 &  & Nasal & 0.0880 & 0.0881 \\
 &  & Temporal & 0.0669 & 0.0688 \\ \midrule
\multirow{7}{*}{\begin{tabular}[c]{@{}c@{}}Train: ORIGA\\ (all images)\\ Test: G1020\\ (all images)\end{tabular}} & \multicolumn{2}{l}{Cup Diameter} & 0.1396 & 0.1031 \\
 & \multicolumn{2}{l}{Disc Diameter} & 0.0593 & 0.0692 \\
 & \multicolumn{2}{l}{CDR} & 0.1674 & 0.1181 \\
 & \multirow{4}{*}{\begin{tabular}[l]{@{}l@{}}Neuroretinal\\ Rim\end{tabular}} & Inferior & 0.2102 & 0.2170 \\
 &  & Superior & 0.2066 & 0.1278 \\
 &  & Nasal & 0.2177 & 0.1933 \\
 &  & Temporal & 0.2150 & 0.1483 \\ \bottomrule
\end{tabular}%
\end{table}
Using correctly predicted OD and OC we then calculate predicted CDR and size of neuroretinal rim in inferior, superior, nasal and temporal quadrants. Mean Absolute Percentage Error (MAPE) between various predicted values and ground truth values is given in Table~\ref{tab:MAPE}. All the values in this table are calculated using IOU$>0.5$.

\subsection{Classification of Glaucoma}
After localising and extracting ODs from the whole fundus images, we used these extracted discs to train inception v3 for classification of healthy and glaucomatous images. We employed 6-fold cross validation with respect to patients to ensure that all images belonging to one patient are in either training set or validation set. The inception model with same experimentation setup was also used to classify ORIGA dataset using 5-fold cross validation. We also evaluated the performance of state-of-the-art method on ORIGA presented by Bajwa et al.~\cite{bajwa2019two} for detection of glaucoma in G1020 dataset. Table~\ref{tab:classResults} shows performance metrics for both classifiers on both datasets. It is evident from the Table that both network were able to classify images from ORIGA with high precision and recall. However, the same networks struggled hard against G1020. We believe that the difference between the performance of inception network on these two datasets is correlated with the way these datasets are collected. ORIGA, and most other publicly available RFI datasets impose many constraints on imaging techniques and selection of images into final dataset that the resulting image set is no longer representative of realistic image capturing practices. A DL model trained on such carefully curated datasets could have the ability to perform well in laboratory conditions but is likely to be unsuccessful in the field.

\begin{table*}[h!]
\centering
\caption{Performance metrics for glaucoma detection on G1020 and ORIGA.}
\label{tab:classResults}
\begin{tabular}{@{}cclccc@{}}
\toprule
\textbf{Method} & \textbf{Dateset} & \multicolumn{1}{c}{\textbf{Class}} & \textbf{Precision} & \textbf{Recall} & \textbf{F1-Score} \\ \midrule
\multirow{6}{*}{\textbf{inception v3}} & \multicolumn{1}{l}{\multirow{3}{*}{\textbf{ORIGA}}} & Healthy & 0.8578$\pm$0.0383 & 0.9170$\pm$0.0208 & 0.8861$\pm$0.0252 \\
 & \multicolumn{1}{l}{} & Glaucoma & 0.6947$\pm$0.0869 & 0.5581$\pm$0.1408 & 0.6157$\pm$0.1165 \\
 & \multicolumn{1}{l}{} & Total & 0.8157$\pm$0.0486 & 0.8246$\pm$0.0419 & 0.8164$\pm$0.0476 \\ \cmidrule(l){2-6} 
 & \multicolumn{1}{l}{\multirow{3}{*}{\textbf{G1020}}} & Healthy & 0.7150$\pm$0.1053 & 0.8183$\pm$0.0289 & 0.7587$\pm$0.0619 \\
 & \multicolumn{1}{l}{} & Glaucoma & 0.2894$\pm$0.0834 & 0.1920$\pm$0.0637 & 0.2219$\pm$0.0513 \\
 & \multicolumn{1}{l}{} & Total & 0.6055$\pm$0.094 & 0.6344$\pm$0.0722 & 0.6080$\pm$0.0988 \\ \midrule
\multirow{6}{*}{\textbf{\begin{tabular}[c]{@{}c@{}}Bajwa et al.\\ (2019)~\cite{bajwa2019two}\end{tabular}}} & \multirow{3}{*}{\textbf{ORIGA}} & Healthy & 0.8231$\pm$0.0288 & 0.9186$\pm$0.0229 & 0.8681$\pm$0.246 \\
 &  & Glaucoma & 0.6552$\pm$0.0665 & 0.4366$\pm$0.0495 & 0.5237$\pm$0.534 \\
 &  & Total & 0.7797$\pm$0.0378 & 0.7938$\pm$0.0342 & 0.7788$\pm$0.0366 \\ \cmidrule(l){2-6} 
 & \multirow{3}{*}{\textbf{G1020}} & Healthy & 0.4735$\pm$0.3348 & 0.6667$\pm$0.4714 & 0.5537$\pm$0.3916 \\
 &  & Glaucoma & 0.0970$\pm$0.1373 & 0.3333$\pm$0.4714 & 0.1503$\pm$0.2126 \\
 &  & Total & 0.0.3646$\pm$0.1979 & 0.0.5706$\pm$0.1976 & 0.4371$\pm$0.2162 \\ \bottomrule
\end{tabular}%
\end{table*}

\subsection{Segmentation of OD and OC}
\begin{figure*}[b!]%
\centering
\begin{subfigure}{\columnwidth}
\includegraphics[width=\columnwidth]{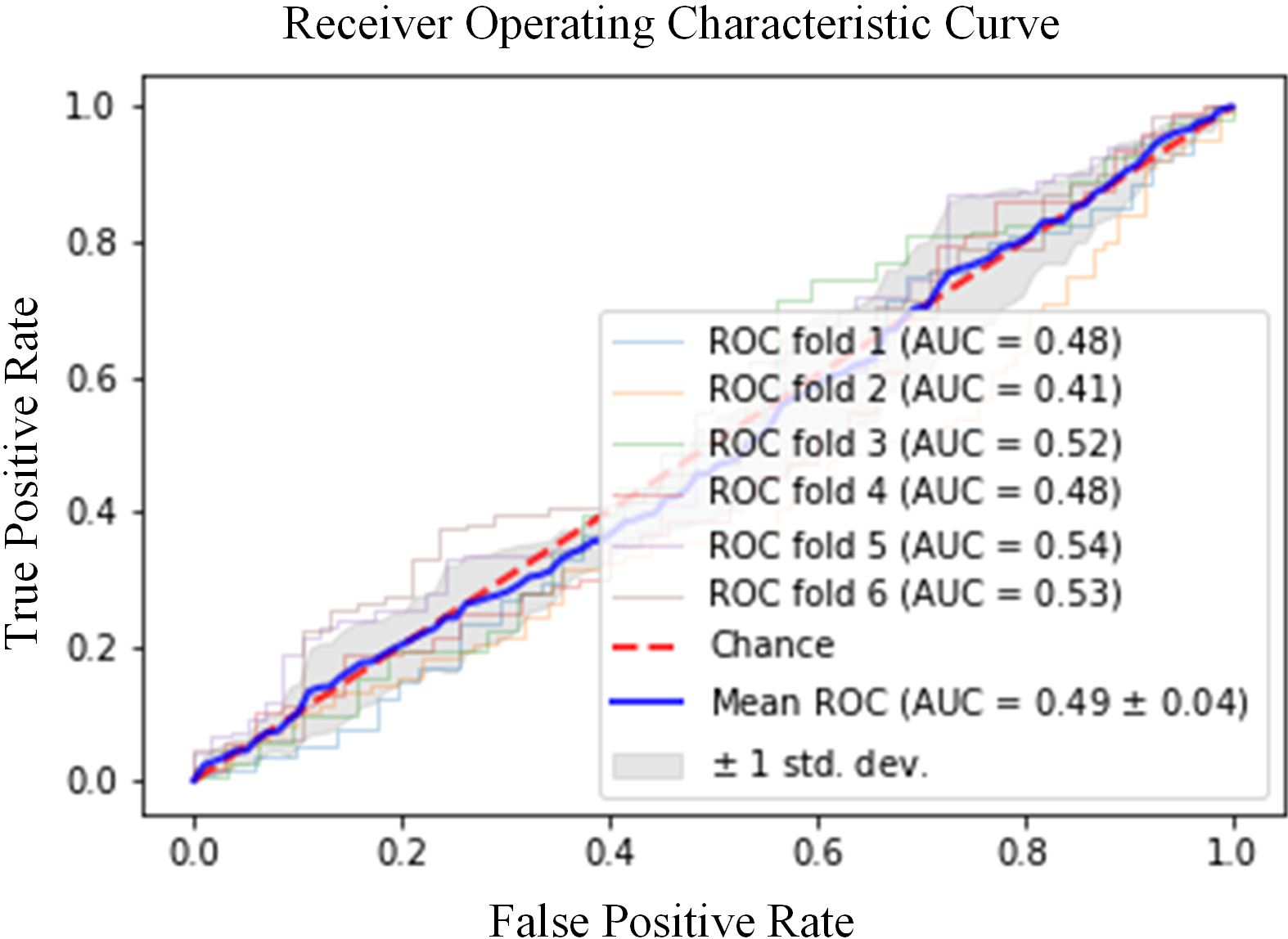}%
\caption{G1020}%
\label{subfig:rocG1020}%
\end{subfigure}
\hfill%
\begin{subfigure}{\columnwidth}
\includegraphics[width=\columnwidth]{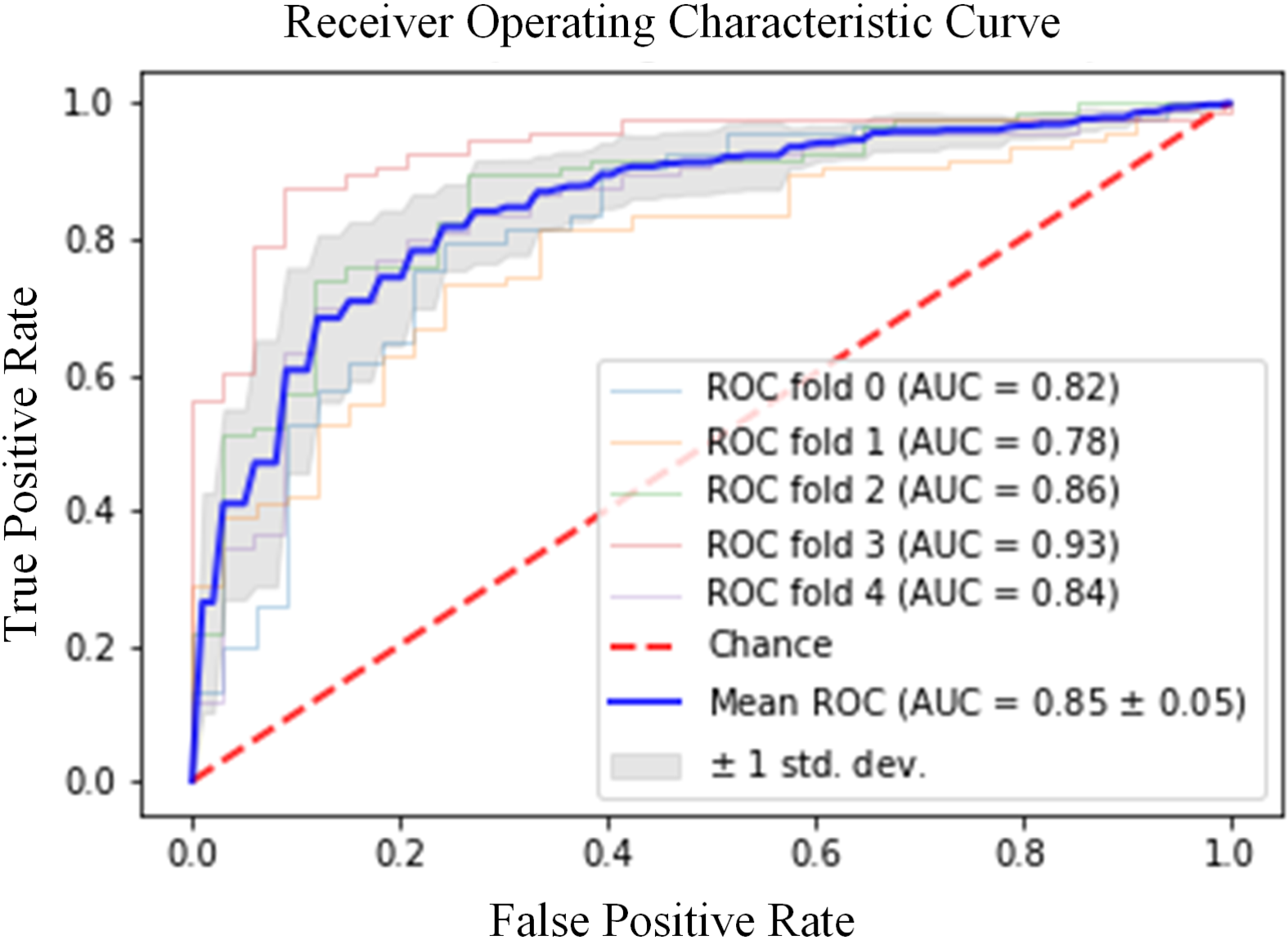}%
\caption{ORIGA}%
\label{subfig:rocORIGA}%
\end{subfigure}
\caption{ROC and AUC for 6-fold G1020 and 5-fold ORIGA datasets.}
\label{fig:ROC}
\end{figure*} 
To provide deeper insight into the complexity of G1020 dataset and compare it with ORIGA, we analysed image embeddings of both datasets from the final convolutional layer of inception model. We applied Principal Component Analysis (PCA) to obtain two of the most significant principal components and visualised them on 2D plane. Fig.~\ref{fig:PCA} illustrates the results of PCA. We can see that glaucoma images (blue dots) and healthy images (red dots) are fairly separable in ORIGA dataset. However, both classes have huge overlap in latent representation of classifier trained on G1020 images.

Fig.~\ref{fig:ROC} shows Area under Receiver Operator Characteristic (ROC) curve for each individual fold and their mean for both datasets. The network was able to achieve competitive AUC compared to state-of-the-art AUC results on ORIGA classification by Bajwa et al.\cite{bajwa2019two} (AUC = 0.874) and Fu et al.~\cite{fu2018joint} (AUC = 0.851), but suffered from serious performance degradation on G1020.

\section{Conclusion}
Most of existing RFI datasets for glaucoma detection are very small in size (a few hundred images) and almost all of them are collected in a very controlled environment. These datasets do not consider practical limitations in imaging and usually exclude images that have other retinal artefacts~\cite{diaz2019cnns}. It has been reported in the literature that presence of multiple eye diseases degrades the performance of DL algorithms trained on such datasets~\cite{al2018dense}. Due to these reasons, most of publicly available datasets for glaucoma detection are unable to train a robust CAD system that can perform equally well in real clinical environment. In this paper, we have presented a new large publicly available dataset of RFIs that closely represents fundus imaging in practical clinical routine and does not enforce strict inclusion criteria on the captured images. Our initial evaluation of various DL methods for OD and OC segmentation and glaucoma classification highlights challenges that need to be addresses to develop a practical CAD system for swift and reliable glaucoma screening. Our results set a baseline for comparison by future works in this domain. We invite research community to utilise this dataset and evaluate their segmentation and classification algorithms on it.

\bibliographystyle{IEEEtran}
\bibliography{references.bib}

\end{document}